# Dynamical amplification of phase conjugation using a modulated optical feedback in a Nd:YVO$_4$ laser

**Hervé Gilles,[1*] Sylvain Girard,[1] Mathieu Laroche,[1] and Eric Lacot[2]**

[1] CIMAP, UMR 6252 CNRS, CEA, ENSICAEN, Université de Caen, ENSI Caen – 6 bd Maréchal Juin, 14050 Caen, France

[2] Laboratoire de Spectrométrie Physique, UMR CNRS 5588, Université Joseph Fourier, 140 Avenue de la physique, BP 87 - 38402 Saint Martin d'Hères,

[*] *Corresponding author:* herve.gilles@ensicaen.fr

We demonstrate an efficient dynamical amplification of phase conjugation in the gain medium of a diode-pumped Nd$^{3+}$:YVO$_4$ solid state laser via excitation of its relaxation oscillations. Consequently, enhancement in the modulated amplitude of the phase conjugate wave is observed with up to +30 dB compared to classical homodyne approach.

Many theoretical and experimental works have already demonstrated that the dynamical behavior of laser can be significantly affected by coherent optical feedback [1][2]. For instance, optical feedback is becoming an efficient optical sensing technique for different applications like telemetry [3], velocimetry [4], vibrational analysis [5] or imaging [6]. This technique is intrinsically very efficient with single frequency class B lasers as they present strong relaxation oscillations after a time dependant perturbation. When submitted to a frequency shifted optical feedback, a beat note between the intracavity oscillating wave and the optical feedback leads to an intensity modulation. When the beating frequency is adjusted close to the relaxation oscillations frequency of the laser, the response to the optical feedback is strongly enhanced. In this scheme, also known as Laser Feedback Interferometry (LFI), the laser cavity plays simultaneously three roles: (i) a coherent light source, (ii) an interferometer and (iii) an optical amplifier. The enhancement factor of the detected interferometric signal is of the order of $10^6$ when the optical beating frequency is resonant with the relaxation oscillations frequency of a solid-state laser. The main advantage of such a technique is therefore to be very sensitive to a very small amount of re-injected light.

Phase conjugation by four-wave mixing (FWM) in saturable solid state amplifying medium has been mainly demonstrated in class B solid-state lasers [7]-[11]. In continuous-wave operation or in pulsed regime, all experiments of FWM have been realized without taking into account the specific dynamical properties of solid state lasers. However, FWM experiments inside a laser cavity can be directly compared to optical feedback experiments. In intracavity FWM experiments, a part of the laser output is used as a signal beam and is re-injected back into the amplifying medium. These experiments look therefore very similar to LFI experiments except that: (1) pump and signal beams are not collinear and (2) no dynamical modulation is applied to the signal wave.

In this Letter, we show that traditional FWM in solid-state laser can be strongly enhanced taking into account the specific dynamical properties of the laser oscillator. For the first time to our knowledge, we show that an amplification effect can be obtained during FWM in a solid state laser via an excitation of its relaxation oscillations. The basic principle consists in a classical FWM experiment except that the intensity of the re-injected signal beam is now amplitude modulated with a sinusoidal signal at a modulation frequency close to the relaxation oscillations frequency of the laser.

The experimental setup is illustrated in figure 1 and is quite similar to the one described in [9]. The diode-pumped $Nd^{3+}$:YVO$_4$ is a simple hemispherical cavity with a cavity length close to 70 mm. The 1% doped Nd:YVO$_4$ crystal is longitudinally pumped at 808 nm using a 4 W fiber-pigtailed laser diode with a spot size around 100 µm. One side of the crystal is antireflection coated at 808 nm and highly reflective at 1064nm and corresponds to the back mirror. The external output coupler has a radius of curvature ROC=80mm with 30% transmission at 1064 nm. In order to keep a TEM$_{00}$ mode, the pump power is limited to 3W giving an output power typically equal to 600mW. The output beam is collimated using a lens L$_1$ with 100mm focal length. In order to avoid any direct optical feedback, an optical isolator (OI) is added just after L$_1$. Compared to the setup described by Brignon et al. [9], an acousto-optic modulator (AOM) is aligned on the optical path of the signal beam for amplitude modulation (AM). The external AM frequency $f$ is adjusted close to the relaxation oscillation frequency of the laser (typically $f = f_{ro} = 1$ MHz) with a modulation depth adjustable from 1% up to 70%. Most of the laser output is therefore used as the signal beam which can be injected back into the Nd:YVO$_4$ medium. As shown in figure 1, a second lens L$_2$ (f=100mm) is used to focus the signal beam into the Nd:YVO$_4$ crystal. Taking into account the losses into the optical isolator and the Bragg cell efficiency, the signal power available for FWM inside the gain medium is close to 250mW whereas the interacting laser beams (A$_{p1(2)}$, A$_s$) inside the crystal have a diameter around 100-150 µm. The angle between the signal beam and the optical axis of the cavity is lower than 5° in order to properly overlap the two counter-propagating pump beams and the signal beam in the gain volume. A beam splitter (BS) is added to monitor on one side the injected signal beam and on the other side the phase conjugate beam. The experimental setup was first aligned without amplitude modulation in order to check evidence of phase conjugation. The conversion efficiency is estimated to 0.07 % quite similar to [9] considering the output coupler. For all the results presented below, only the modulated amplitude of the phase conjugate beam is measured.

It is well known that FWM in a laser cavity results from the interference pattern created inside the amplifying medium by the two counter propagating pump beams (A$_{p1}$ and A$_{p2}$ in figure 1) interfering with the signal beam A$_s$. The phase conjugate beam A$_c$ results simultaneously from the gain gratings acting in reflection for A$_{p2}$ or transmission for A$_{p1}$ with similar efficiency. We can therefore approximate the phase conjugate intensity as being proportional to:

$$|A_c|^2 \propto \gamma(f) \times R_c(\eta) \times |A_s|^2 \quad (1)$$

with $\quad \gamma(f) = \dfrac{\gamma_c \cdot \sqrt{\dfrac{(\eta \cdot \gamma_1)^2}{4\pi^2} + f^2}}{\sqrt{4\pi^2 \left(f_{ro}^2 - f^2\right)^2 + (\eta \cdot \gamma_1)^2 f^2}} \quad$ and $\quad f_{ro} = \dfrac{1}{2\pi}\sqrt{\gamma_1 \gamma_c (\eta - 1)} \quad (2)$

$R_c$ is calculated thanks to equation (9) of ref. [13] and depends on $A_{p1(2)}$ and $\eta$ the normalized pumping (808 nm) rate. In order to take into account the effect of AM applied on the signal $A_s$, the amplification factor $\gamma(f)$ defined in [12] is used in which $\gamma_1$ (4000 s$^{-1}$) is the decay rate of the population inversion, $\gamma_c$ (4.10$^9$ s$^{-1}$) is the laser cavity decay rate, $f_{ro}$ is the relaxation oscillations frequency of the laser and $f$ the AM frequency.

The factor $\gamma(f)$ shows a strong resonance near the laser relaxation oscillations frequency. In order to check if such enhancement could also be obtained in FWM, the influence of the AM frequency $f$ applied to the signal beam is investigated. When the AM frequency is far away from the relaxation oscillations frequency of the laser, the conjugate beam presents a very small intensity modulation depth. On the other hand, the phase conjugate wave exhibits a strong resonance when the AM frequency is adjusted close to the relaxation oscillations frequency. When the incident modulation depth is higher than 30%, a periodic pulsed regime with a repetition rate equal to the AM frequency and a pulse duration of 250ns was observed. At smaller incident modulation depth ($\leq 1\%$), the temporal evolution of the phase conjugate intensity becomes a purely sinusoidal waveform but still presenting a strong resonance when $f = f_{ro}$. When the signal beam intensity is modulated, it creates in the gain medium an inversion population grating which is also amplitude modulated at the same frequency. The resulting modulation in the amplifying medium allows exciting the relaxation oscillations. In this case, the time dependant perturbation is strongly amplified inside the laser cavity. For a better characterization of the phase conjugate wave, figure 2 shows the evolution of the phase conjugate wave intensity ($|A_c|^2$) versus $f$ superimposed with the random intensity noise (RIN) spectrum of the unperturbed laser. This experiment has been realized for a low modulation depth (0.5%). It clearly shows that an amplification factor up to +30 dB appears when $f = f_{ro}$. The evolution shown in figure 2 is directly proportional to the $\gamma(f)$ function defined in equation (2) and illustrates the exaltation effect provided by the specific dynamical properties of a class B solid state laser. The evolution of FWM in a class B laser submitted to a modulated optical feedback which is slightly titled seems to be quite similar to the one already observed in self-aligned optical feedback experiments.

Finally, the phase conjugate wave intensity is investigated versus the optical power of the signal beam. In a first experiment, only the signal beam power is attenuated using a continuously adjustable neutral density filter located just before the AOM. The intensities $|A_{p1}|^2, |A_{p2}|^2$ and $\gamma(f)$ are kept constant. The different curves reported in figure 3(a) are obtained for three different values of the AM frequency $f$. As expected from equation (1), $|A_c|^2$ is linearly proportional to the signal power $|A_s|^2$. An amplification effect – which corresponds to an increase in the slope efficiency of the curves reported in figure 3 – is effectively observed when the modulation frequency is very close to the relaxation oscillations frequency of the laser. Similar measurements were performed when the power from the diode at λ=808 nm was adjusted meaning that the pumping rate η was adjusted from 1.3 to 3.5. In this case, the pump and signal powers ($|A_{p1}|^2, |A_{p2}|^2, |A_s|^2$) and $\gamma(f)$ are simultaneously affected. Solid lines in figure 3b take into account these simultaneous variations. Figures 3(a) and 3(b) are in good agreement with equations (1) and (2).

To conclude, we have experimentally demonstrated an amplification effect of intracavity FWM in a diode pumped Nd:YVO$_4$ laser using a selected excitation of the relaxation oscillations of the laser. A +30 dB amplification factor has been measured on the modulated amplitude of the phase conjugate beam when the AM frequency is resonant with the relaxation oscillations frequency of the laser. Moreover, we plan to demonstrate full-field imaging capabilities of such a device.

Figures captions

Figure 1 : Experimental setup of a modulated FWM experiment in a diode–pumped Nd:YVO$_4$ laser; M, high-reflectivity plane mirrors; C, Nd:YVO4 crystal; M$_o$, laser output coupler; AOM, acousto-optic modulator; OI, optical isolator; D$_{1,2}$, InGaAs photodiodes; L$_{1,2}$, lens; BS, beam-splitter.

Figure 2 : Evolution of the amplitude modulated part of the conjugate wave intensity versus the AM frequency compared to the noise spectrum of the free-running laser driven by white noise;

Figure 3 : Evolution of the amplitude modulated part of the conjugate wave intensity versus the signal beam power; (a) only the signal beam power $|A_s|^2$ is modified; (b) $|A_{p1}|^2$, $|A_{p2}|^2$, $|A_s|^2$ are modified in the same way.

Figure 1

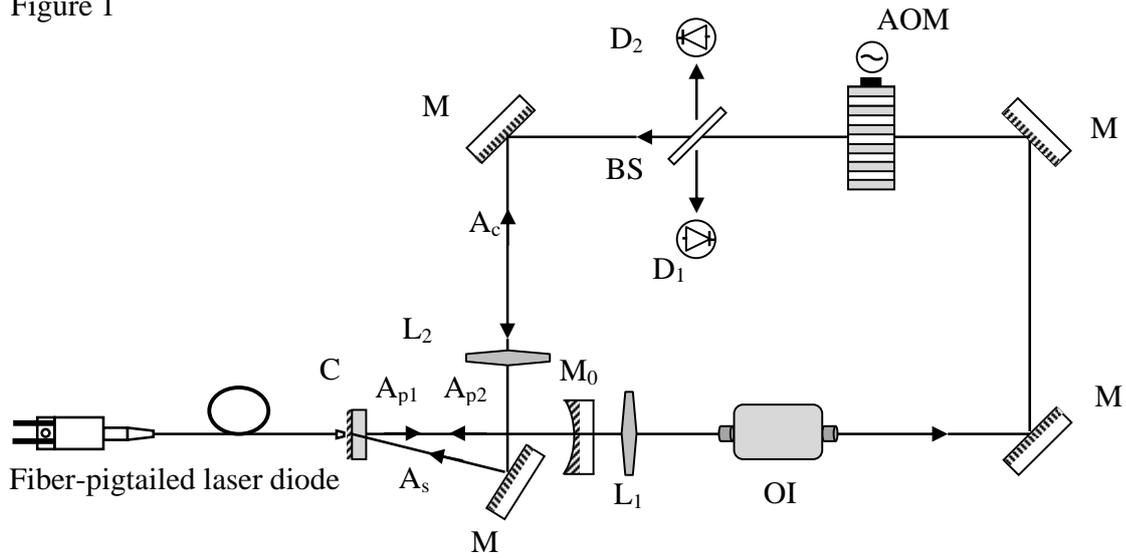

Figure 2

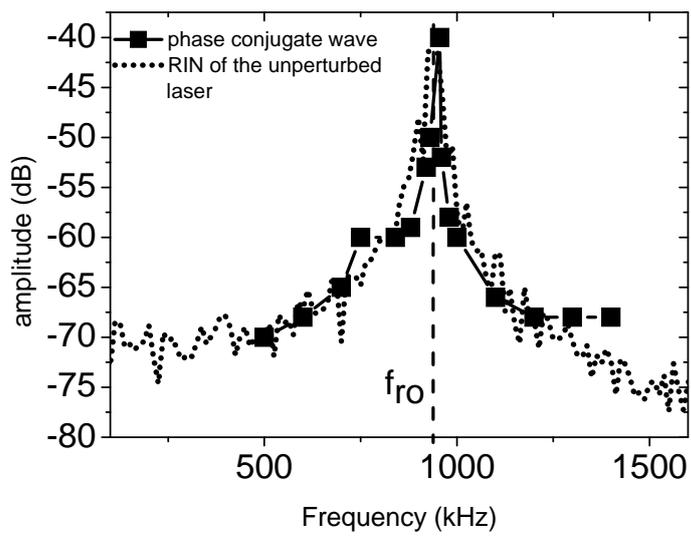

Figure 3

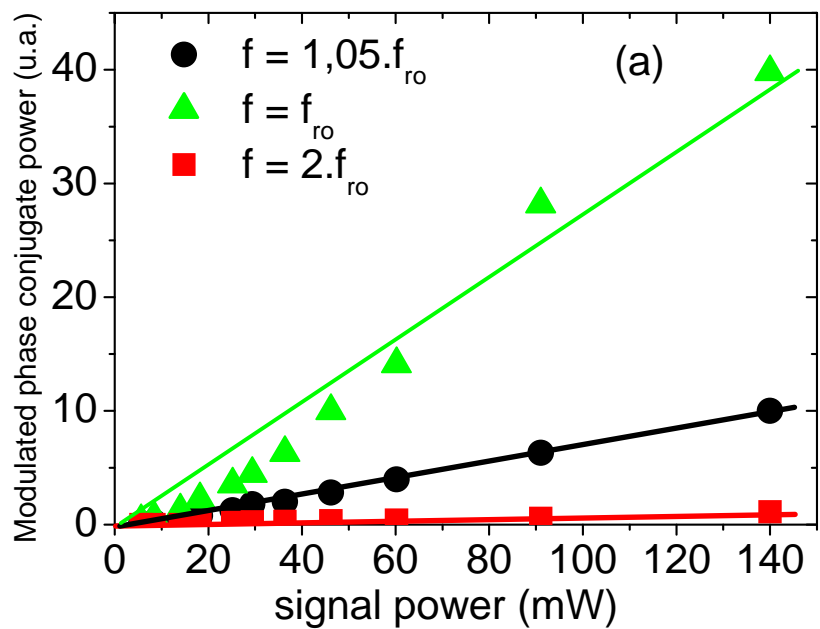

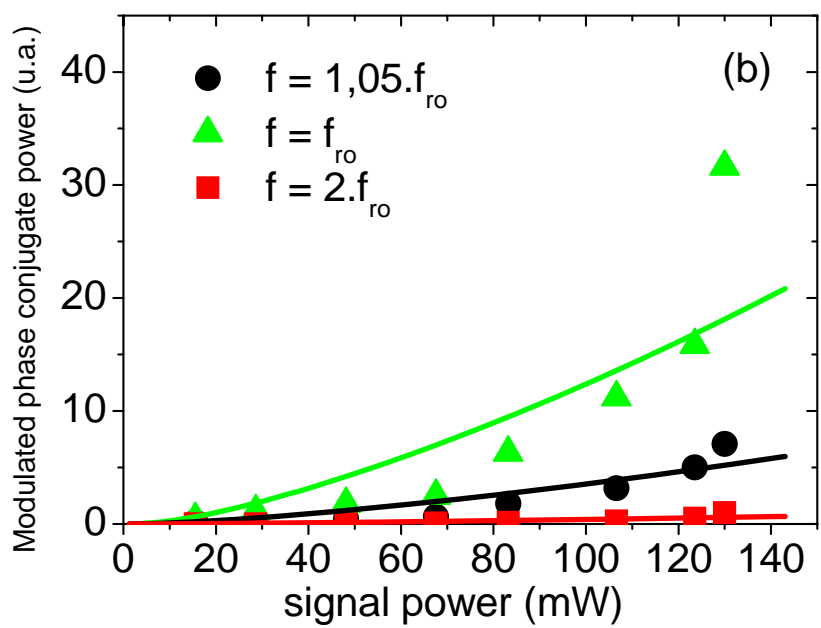